\documentclass[aps,prx,reprint,superscriptaddress]{revtex4-2}

\usepackage{graphicx}  

\usepackage{longtable}
\usepackage[T1]{fontenc}
\usepackage{dcolumn}   
\usepackage[inline]{enumitem}

\usepackage{bm}        
\usepackage{amsfonts}  
\usepackage{amsmath}   
\usepackage{amssymb}   
\usepackage{color}
\usepackage[colorlinks=true,allcolors=blue]{hyperref}

\newcommand{\pwisein}{\left\{ \begin{array}{ll}}
\newcommand{\pwiseout}{\end{array}\right.}

\begin{document}

\title{Stochastic thermodynamics of a quantum dot coupled to a finite-size reservoir}

\author{Saulo V. Moreira}
\affiliation {Department of Physics and NanoLund,  Lund University, Box 118, 22100 Lund, Sweden. }
\affiliation {School of Physics, Trinity College Dublin, Dublin 2, Ireland. }

\author{Peter Samuelsson }
\affiliation {Department of Physics and NanoLund,  Lund University, Box 118, 22100 Lund, Sweden. }

\author{Patrick P. Potts}
\affiliation {Department of Physics and Swiss Nanoscience Institute, University of Basel, Klingelbergstrasse 82, 4056 Basel, Switzerland.}

\begin{abstract}  

In nano-scale systems coupled to finite-size reservoirs, the reservoir temperature may fluctuate due to heat exchange between the system and the reservoirs. To date, a stochastic thermodynamic analysis of heat, work and entropy production in such systems is however missing. Here we fill this gap by analyzing a single-level quantum dot tunnel coupled to a finite-size electronic reservoir. The system dynamics is described by a Markovian master equation, depending on the fluctuating temperature of the reservoir. Based on a fluctuation theorem, we identify the appropriate entropy production that results in a thermodynamically consistent statistical description. We illustrate our results by analyzing the work production for a finite-size reservoir Szilard engine.

\end{abstract}


\maketitle 



{\it Introduction.}--- In nanometer scale systems in contact with an environment, fluctuations of physical quantities are ubiquitous. The ability to control and measure systems at such small scales has been a key driving force in the development of stochastic thermodynamics~\cite{Harris2007, Esposito2009, Jarzynski2011, Seifert2012,Bochkov2013,vandenBroeck2015,Mansour2017,Seifert2019}, which provides a theoretical framework for thermodynamics phenomena based on concepts such as stochastic entropy~\cite{Seifert2005}, as well as detailed~\cite{Bochkov1981,Bochkov1981b,Crooks1998,Crooks1999,Crooks2000,Jarzynski2000,Seifert2004} and integral~\cite{Jarzynski1997, Jarzynski1997PRE, Rastegin2013} fluctuation theorems. Stochastic thermodynamics has over the last two decades successfully been employed to describe a large number of experiments on small scale systems, such as implementations of Maxwell’s demon~\cite{Toyabe2010, KoskiPRL2014, Vidrighin2016, Camati2016, Cottet2017} and Szilard’s engine~\cite{Koski2014, Barker2022}, verifications of Landauer’s principle~\cite{Serreli2007, Berut2015, Hong2016}, tests of fluctuation theorems~\cite{Liphardt2002, Collin2005, Saira2012, Hoang2018}, and determination of system free energies~\cite{Hummer2001, Collin2005, Hummer2010, Watanabe2023}. 

In all these experiments, the environment can to a good approximation be described as a bath, or reservoir, in thermal equilibrium. Thus, the reservoir is effectively of infinite size, such that the exchange of heat with the system does not affect the reservoir.  However, in many nanoscale experiments, the reservoirs are themselves of finite sizes, with system back action inducing energy fluctuations  within the reservoir ~\cite{Pekola2021, Dutta2020, Jackson2022, Spiecker2023, SpieckerPavlov2023, Champain2023}. Given a fast relaxation time-scale, such a reservoir may then be described by a fluctuating temperature. Such temperature fluctuations were recently investigated in small metallic islands~\cite{Karimi2020}. 

Theoretically, the effect of finite-size reservoirs with time-dependent (but not fluctuating) temperatures on thermodynamic and transport properties have been investigated in a number of systems~\cite{Gallego2014,Schaller2014, Grenier2016,Amato2020, Ma2020, Yuan2022}. Furthermore, average values of thermodynamic quantities have been investigated for finite-size reservoirs that exhibit energy fluctuations~\cite{RieraCampeny2021}. There is to date, however, no stochastic thermodynamics analysis of small scale systems coupled to finite-size reservoirs, fully accounting for the system-reservoir back action and the resulting, correlated fluctuations of their physical properties. While the formalism outlined in Ref.~\cite{Strasberg2021-1} could provide the basis of such an investigation, we focus here on scenarios where the reservoir may be described by a (fluctuating) temperature at all times.

In this letter, we present such a stochastic thermodynamics analysis, focusing on a basic, experimentally realizable setup~\cite{Dutta2020, Champain2023} – a single level quantum dot with a time-dependent level energy, tunnel-coupled to a finite-size electronic reservoir that can be described by a fluctuating temperature. The dynamics of the system and the reservoir temperature is described by a Markovian master equation. Based on a fluctuation theorem, relating the probabilities for forward and backward trajectories for the system and the reservoir temperature,  we identify the appropriate stochastic entropy production. This allows for a thermodynamically consistent description given the knowledge of the fluctuating reservoir temperature. This is in contrast to previous approaches describing finite-size reservoirs, where effective temperatures are defined based on averages of reservoir observables~\cite{Strasberg2021-1, Strasberg2021-2, Elouard2022}.  To illustrate the approach, we consider a Szilard engine and show that the performed work is smaller than the work of an ideal engine, where the reservoir is of infinite size.

\begin{figure}

\includegraphics[width=3.3in]{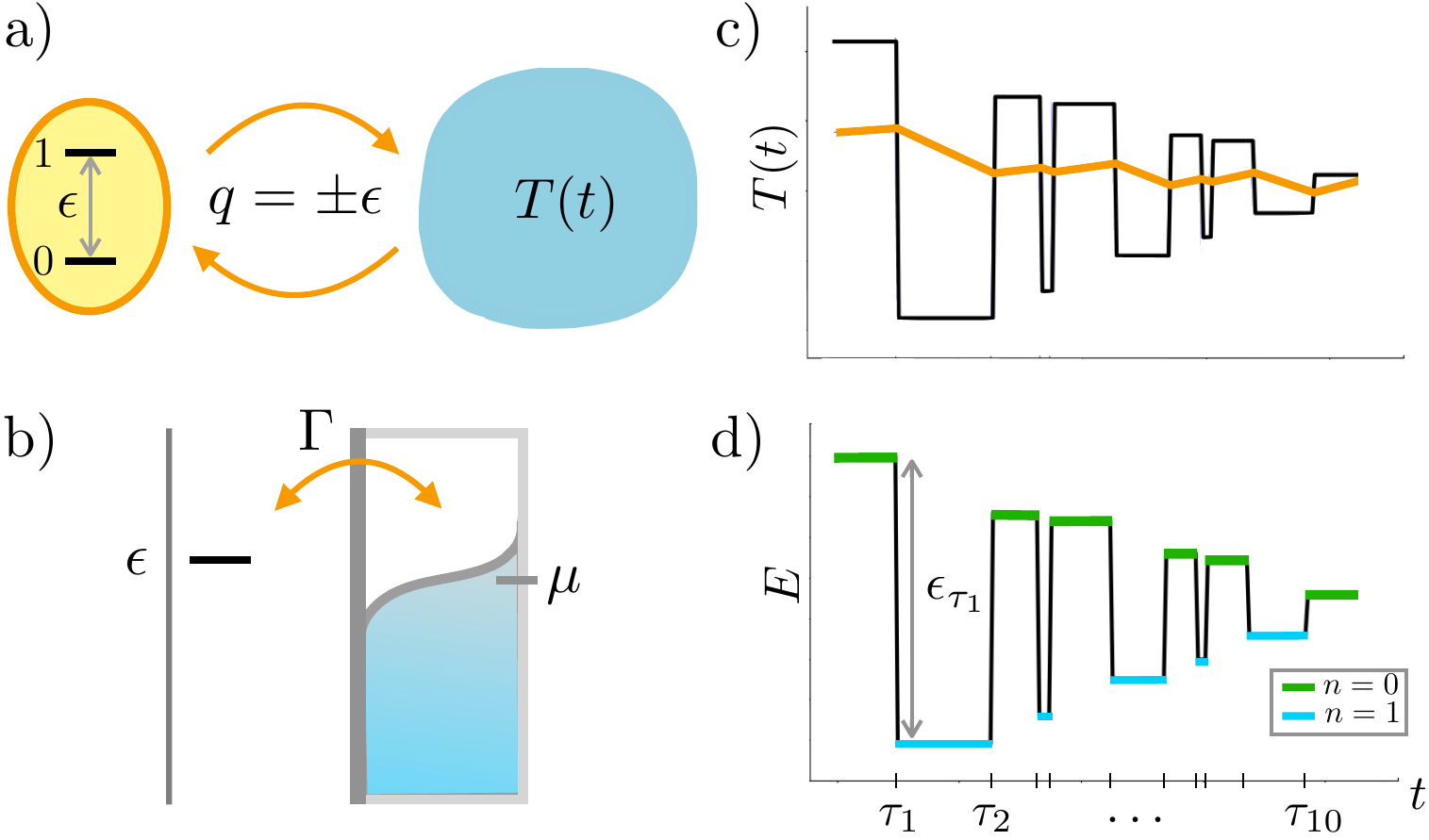}
\caption{\label{setup} 
a) Sketch of a two-level system with an energy gap $\epsilon$, coupled to a finite-size reservoir, the temperature of which being $T(t) \equiv T(E(t))$. The system and reservoir exchange discrete amounts of heat $q=\pm \epsilon$ in a stochastic way.
b) Representation of the coupling between the dot system and the finite-size fermionic reservoir, where $\Gamma$ is the tunneling strength.
c) Plot of the temperature $T(t)$ and the entropic temperature $T_{\text e }(t)$ in orange for a linearly decreasing level energy $\epsilon_t$.
d) Reservoir energy as a function of time. When an electron tunnels at time $\tau_j$, the reservoir energy changes by $\epsilon_{\tau_j}$.
} 

\end{figure}

{\it Entropy production conundrum.}---The challenges with a stochastic thermodynamics description of small systems coupled to finite-size reservoirs can be compellingly illustrated by considering the basic setup in Fig. \ref{setup}\,a). 
A classical two-state system, with an energy difference $\epsilon>0$ between the two states 0 and 1, exhibits stochastic state transfers due to the exchange of a discrete amount of heat $q=\pm \epsilon$ with a finite-size reservoir. Assuming that the reservoir temperature $T$ increases monotonically with increasing reservoir energy, $T$ will fluctuate between $T_0$ and $T_1<T_0$, with superscripts denoting the system state. Naively employing the known result for infinite-size reservoirs, namely that the entropy is given by the heat transferred divided by the reservoir temperature, one would assume that a transfer of heat into (out of) the reservoir leads to a production of entropy $\Delta s_{\text{in}}=\epsilon/T_1$ ($\Delta s_{\text{out}}=-\epsilon/T_0$) in the reservoir. For two subsequent heat transfers this leads to a reservoir entropy production
\begin{equation}\label{StochEntr}
\Delta s_{\text{in}}+\Delta s_{\text{out}}=\epsilon(1/T_1-1/T_0)>0
\end{equation}
Since the system is back in the same state after two transfers, the system does not contribute to the entropy. The result in Eq.~\eqref{StochEntr} would thus imply that in equilibrium, stochastic heat fluctuations lead to a non-zero entropy production, which is physically non-sensical. 
From this reasoning, it is clear that entropy production in a reservoir with a temperature changing as a result of heat transfers between system and reservoir requires further understanding. To provide this, in the following we present a fully stochastic approach to the thermodynamics of an experimentally realistic implementation of the setup in Fig. \ref{setup}\,a).

{\it System and master equation.}---We consider a single level quantum dot with a time-dependent level energy $\epsilon_t$, coupled to a finite-size electron reservoir via a tunnel barrier, characterized by a tunnel rate $\Gamma$, see Fig.~\ref{setup}\,b).
An electron tunneling from the dot to the reservoir (from the reservoir to the dot) at time $t$ adds (removes) energy $\epsilon_t$ to (from) the reservoir.
The stochastic nature of the tunneling process induces, in this way, fluctuations in time of the reservoir energy $E$.
The electronic thermalization in the reservoir is considered to be so fast that, at all times, the electrons are effectively in a quasi-equilibrium state, described by a Fermi distribution
\begin{equation}\label{fermi}
f(\epsilon,E)=\frac{1}{1+e^{\epsilon/k_\text B T(E)}}.
\end{equation}
The temperature $T(E)$ is related to $E$ via the heat capacity $C(T)=C'T$ as
\begin{equation}\label{Temperature}
 T(E) = \sqrt{\frac{2E}{C^\prime}},
\end{equation}
where $C^\prime = \pi^2 k_\text B^2 \nu_0/3$, $\nu_0$ being the density of states and $k_\text B$ is the Boltzmann constant. As $E$ fluctuates in time, $T(E)$ is a fluctuating temperature.



We describe the system's time evolution with the phenomenological, energy resolved master equation,
\begin{equation}\label{MasterEquation}
\begin{aligned}
  &\dfrac{d}{dt}  \begin{bmatrix}  p_0(E) \\  p_1(E-\epsilon_t) \end{bmatrix}  =  \mathcal{W} \begin{bmatrix}  p_0(E) \\  p_1(E-\epsilon_t) \end{bmatrix},\\
  &\mathcal{W}=\begin{bmatrix}  -\Gamma_{\text{in}}(\epsilon_t, E)  &\Gamma_{\text{out}}(\epsilon_t,E-\epsilon_t) \\  \Gamma_{\text{in}}(\epsilon_t,E) &-\Gamma_{\text{out}}(\epsilon_t,E-\epsilon_t) \end{bmatrix}
  \end{aligned}
\end{equation}
where $p_n(E)\equiv p(n,E;t)$ is the probability that there are $n=0,1$ electrons on the dot and that the reservoir energy is $E$ at time $t$. The probabilities satisfy the normalization condition  $\int dE\ [p_0(E) + p_1(E)] = 1$, and the tunneling rates are given by~\cite{Golovach2011,vandenBerg2015}
\begin{align}\label{TunnelingRateOUT}
    \Gamma_{\text{in}}(\epsilon, E) = \Gamma f(\epsilon, E), \ \
    \Gamma_{\text{out}}(\epsilon, E) = \Gamma[1 - f(\epsilon, E) ], 
\end{align} 
see the supplemental information (SI) for a motivation of these rates, as well as a discussion on charging effects and chemical potential fluctuations.

For a level energy that is constant in time, $\epsilon_t = \epsilon$, given a well-defined initial {\it total} energy for the system and reservoir (denoted by $\mathcal{E}$), the reservoir energy can only take on the values $\mathcal{E}$ and $ \mathcal{E} - \epsilon$, for zero and one electron on the dot, respectively.
This implies that the temperature fluctuates between $T(\mathcal{E}) $ and $T(\mathcal{E}-\epsilon) $, and the stationary solution to Eq.~\eqref{MasterEquation} is given by $p_n(E)=\delta(E-\mathcal{E}+n\epsilon)p^{\text s}_n(\epsilon|\mathcal{E})$ with
\begin{equation}\label{Stationary}
p_1^{\text s}(\epsilon|\mathcal{E}) = 1 - p_0^{\text s}(\epsilon|\mathcal{E}) = \frac{f(\epsilon, \mathcal{E})}{1 - f(\epsilon, \mathcal{E}-\epsilon) + f(\epsilon, \mathcal{E})},
\end{equation}
which reduces to $f(\epsilon,\mathcal{E})$ only when $T(\mathcal{E}-\epsilon)\simeq T(\mathcal{E})$.




{\it Fluctuation theorem and entropic temperature}.--- For a stochastic thermodynamic description, we consider $n(t)$ and $E(t)$ as the stochastic system state and reservoir energy, respectively. A trajectory $\gamma = \{n(t),E(t)|0\leq t\leq \tau\}$ may then be defined during a protocol, where the level energy may depend on time $\epsilon_t$, as illustrated in Fig.~\ref{setup}\,d). 
We denote the starting point of $\gamma$ as $(n_0,E_0)\equiv (n(0),E(0))$ and its endpoint as $(n_\tau,E_\tau)\equiv(n(\tau),E(\tau))$.
Note that $E(t)$ and $n(t)$ undergo abrupt changes at times $\tau_j$, whenever an electron tunnels.

A fluctuation theorem relates the probability density $P(\gamma)$ for the trajectory to occur to the probability density $\tilde{P}(\tilde \gamma)$ for the time-reversed trajectory $\tilde \gamma$ to occur under the time reversed protocol (where the level energy is changed as $\epsilon_{\tau-t}$)
\begin{equation}\label{FluctTheorem}
\dfrac{P(\gamma)}{\tilde{P}(\tilde{\gamma})} = \exp{\left[\dfrac{\sigma(\gamma)}{k_\text B}\right]},
\end{equation}
where $\sigma(\gamma)$ is the total, stochastic entropy production along $\gamma$. We can write 
\begin{equation}\label{StochasticEnt}
    \sigma(\gamma) = \Delta s(\gamma) + \Delta s_\text r(\gamma),
\end{equation}
where $\Delta s$, the change in system entropy, is given by
\begin{equation}\label{EntrSystem}
    \Delta s(\gamma) \equiv k_\text B[\ln p(n_0, E_0;0) -\ln p(n_\tau, E_\tau;\tau)],
\end{equation}
where $p(n_0, E_0;0)$, $p(n_\tau, E_\tau;\tau)$ are the probabilities for the initial and final system states and reservoir energies. The term $\Delta s_{\rm r}$, describing the stochastic entropy production associated to the reservoir, can be written as a stochastic integral along the trajectory $\gamma$ (see the SI)
\begin{equation}\label{EntrReservoir}
    \Delta s_\text r(\gamma) \equiv -\int_\gamma \frac{dq(t)}{T_\text{e}(t)},
\end{equation}
where $dq(t)= \epsilon_{ t} dn(t)$ and we introduced the {\it entropic temperature} as
\begin{equation}\label{EffectiveT}
    T_{\text{e}}(t) \equiv T_{\text{e}}(\epsilon_t,\mathcal{E}(t))= \frac{\epsilon_t}{k_\text B} \left[\ln \frac{\Gamma_{\text{out}}(\epsilon_t,\mathcal{E}(t)-\epsilon_t)}{\Gamma_{\text{in}}(\epsilon_t,\mathcal{E}(t))}\right]^{-1},
\end{equation} 
where $\mathcal{E}(t) = E(t) +\epsilon_t n(t)$ denotes the total energy. Note that the entropic temperature is a stochastic variable taking on different values along different trajectories, just like $n(t)$ and $E(t)$.

The entropic temperature in Eq.~\eqref{EntrReservoir} determines how the reservoir stochastic entropy changes along a given trajectory.
In Fig.~\ref{setup}\,c), we illustrate its behaviour in comparison to the actual temperature, $T(t) \equiv T(E(t))$ which is obtained from the stochastic energy $E(t)$ via Eq.~\eqref{Temperature}.
We note that the entropic temperature is a continuous function with kinks when quanta of energy are exchanged via the tunneling process. This is due to the fact that the change in total energy is determined by the work performed on the system, which exhibits kinks because work is only performed when the dot is occupied (see below).

The entropic temperature becomes particularly simple for a constant level energy $\epsilon_t = \epsilon$ (again, assuming a fixed total energy)
\begin{equation}\label{EntrTempS}
    T_{\text{e}}(\epsilon,\mathcal{E}) =  \frac{\epsilon}{k_\text B} \left[\ln \frac{p_0^{\text s}(\epsilon|\mathcal{E})}{p_1^{\text s}(\epsilon|\mathcal{E})}\right]^{-1},
\end{equation}
where the entropic temperature is no longer a stochastic quantity. 
Furthermore, the stationary solution in Eq.~\eqref{Stationary} can be written as
\begin{equation}\label{StationaryProbFermiEff}
    p_1^{\text s}(\epsilon|\mathcal{E}) =\frac{1}{1+e^{\epsilon/k_\text B T_e(\epsilon,\mathcal{E})}}.
\end{equation}
The steady-state occupation of the dot is thus given by the Fermi-Dirac distribution if the entropic temperature is used.
These observations further illustrate that it is the entropic temperature that determines the thermodynamics of the dot coupled to a finite-size reservoir.

{\it Stochastic thermodynamics.}--- The stochastic internal energy of the system along the trajectory $\gamma$ can be defined as
\begin{equation}\label{StochIntEnergy}
    u(t) \equiv n(t)\epsilon_t.
\end{equation}
The average internal energy is obtained by averaging this expression over the distribution for trajectories $P(\gamma)$,
\begin{equation}\label{IntEnergy}
    U(t) = \langle n(t) \rangle\epsilon_t = p_1(t)\epsilon_t,
\end{equation}
where $p_n(t) = \int dE p(n,E;t)$.
According to the first law of thermodynamics, the system's internal energy changes can be divided into work and heat.
Using Eq.~\eqref{StochIntEnergy}, we identify heat and work as
\begin{equation}\label{Stoch1stLaw}
    du(t) = \epsilon_t dn(t)+n(t)\dot{\epsilon}_t dt = dq(t)+ dw(t),
\end{equation}
where the dot denotes a derivative with respect to $t$.
In this way, the stochastic heat and work along the trajectory are given by
\begin{equation}\label{StochWorkHeat}
    q \equiv \int_\gamma  \epsilon_t dn(t), \hspace{.3cm} w \equiv \int_0^{\tau} dt \ n(t) \dot{\epsilon}_t.
\end{equation}
Similarly to Eq.~\eqref{IntEnergy}, we can write the first law in terms of the average heat, $Q\equiv\langle q\rangle$, and average work, $W\equiv\langle w\rangle$,
\begin{equation}\label{1stLaw}
    \Delta{U}\equiv U(\tau)-U(0) = {W} + {Q},
\end{equation}
where
\begin{equation}
Q =  \int_0^{\tau} dt \dot{p}_1(t)\epsilon_t,\hspace{.5cm} \label{Heat} W  = \int_0^{\tau} dt p_1(t)\dot{\epsilon}_t.
\end{equation}

Moreover, we can obtain the average entropy production by averaging the stochastic entropy production in Eq.~\eqref{StochasticEnt}
\begin{equation}\label{EnsembleEntropyProduction}
    \Sigma \equiv \langle \sigma(\gamma)\rangle = \Delta S - \left\langle  \int_\gamma \frac{dq(t)}{T_\text{e}(t)} \right\rangle,
\end{equation}
where $\Delta S \equiv \langle \Delta s(\gamma)\rangle $.
Using Eq.~\eqref{FluctTheorem}, the non-negativity of the  Kullback-Leibler divergence~\cite{Kullback1951} implies that
\begin{equation}\label{SecondLaw}
    \Sigma = k_{\rm B}\left\langle\ln\frac{P(\gamma)}{\tilde{P}(\tilde{\gamma})} \right\rangle\ge 0.
\end{equation}
Hence, Eq.~\eqref{SecondLaw} can be seen as a second law of thermodynamics for the dot system coupled to the finite-size reservoir.
Note that, by considering the entropy production in Eq.~\eqref{EnsembleEntropyProduction} for a time-independent level energy $\epsilon_t = \epsilon$, it follows that $\Sigma = 0$ in equilibrium, as expected. Remarkably, we show in the SI that not only the average entropy production but also the stochastic entropy production in Eq.~\eqref{StochasticEnt} is zero in equilibrium as well as in the quasi-static limit, where the system always approximately remains in equilibrium.

We note that Eqs.~\eqref{EnsembleEntropyProduction} and \eqref{SecondLaw} look just like Clausius' second law~\cite{Clausius1865}, but with stochastic quantities and with temperature being replaced by the entropic temperature. The reason that the entropic and not the actual temperature enters the entropy production is because energy exchange happens in quanta. 
Indeed, we find that for $\epsilon_t \ll \mathcal{E}(t)$
\begin{equation}
    T_{\text{e}}(t) \approx \sqrt{\frac{2\mathcal{E}(t)}{C^\prime}} -  \sqrt{\frac{2\mathcal{E}(t)}{C^\prime}} \frac{\epsilon_t}{4\mathcal{E}(t)}.
\end{equation}
Thus, when the quantization of energy becomes negligible (and $\mathcal{E}(t)\approx E(t)$), the entropic temperature reduces to the actual reservoir temperature $T_\text e (t) \approx T(t)$.
In turn, a sizeable $\epsilon_t$ leads to the disparity between the entropic temperature and the actual temperature. Our equations may thus be understood as a generalization of Clausius' second law that takes into account energy quantization in the exchange of heat.


In the SI, we consider previously derived expressions for entropy production~\cite{Esposito2010,Strasberg2021-2} which are expected to be positive in our scenario.

\begin{figure}
\includegraphics[width=3.5in]{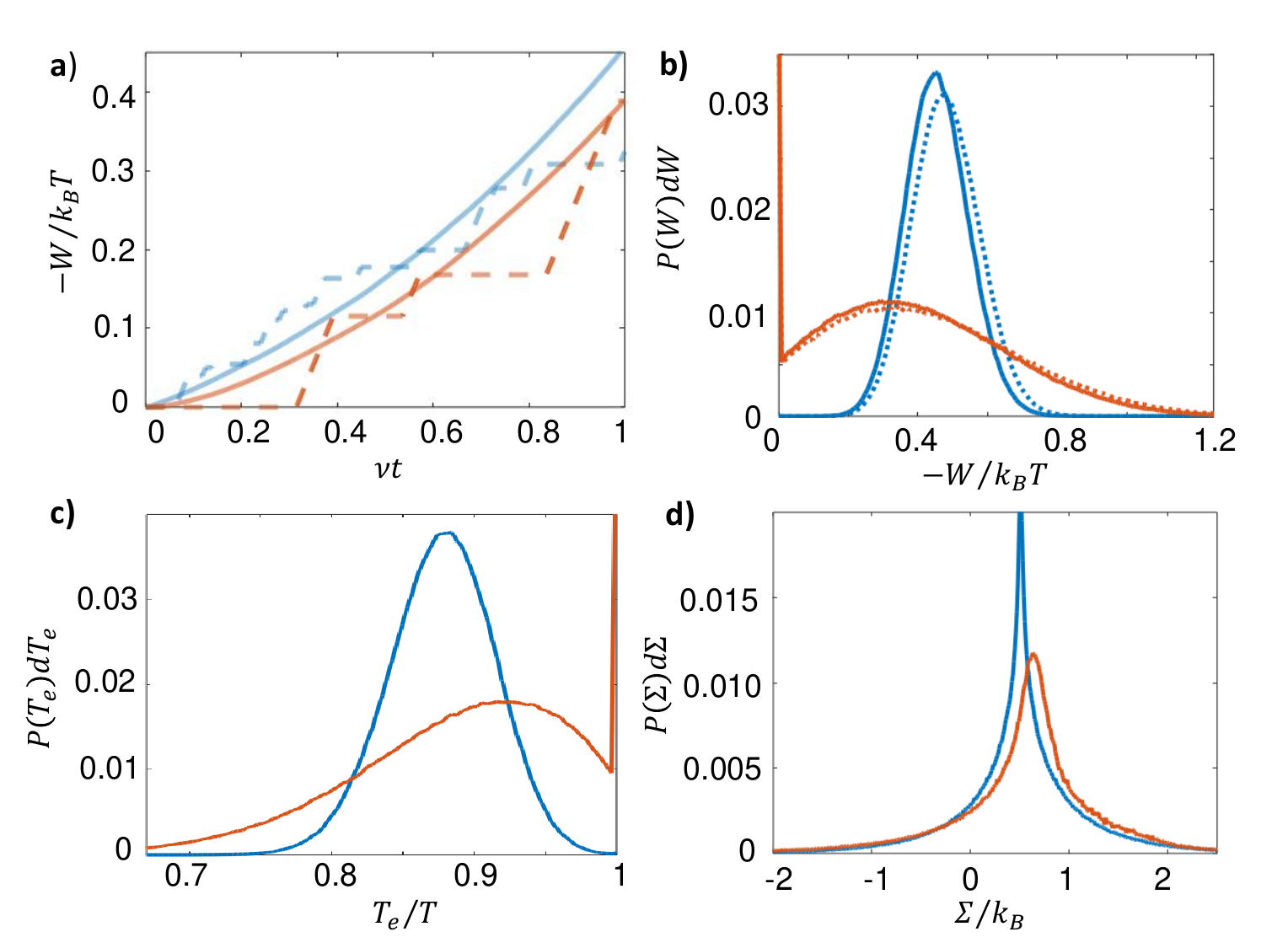}
\caption{\label{Fig2} 
Stochastic thermodynamic quantities. a) The average extracted work $-W$ (solid lines) and the extracted work along a typical trajectory (dashed lines) as a function of time $t$. b) The probability distribution of work extracted during the protocol. Solid (dashed) lines are for a finite (infinite) size reservoir. c) The probability distribution of the entropic temperature $T_e$ at the end of the protocol. d) The probability distributions of the total entropy productions $\Sigma$ during the protocol. In all panels the heat capacity $C(T)=4k_\text B$ and the dot level energy is driven as $\epsilon_t=\epsilon_0(1-\nu t)$ with $\epsilon_0/k_{\rm B}T=1.5$ and $10^6$ trajectories have been generated. Initially, at $t=0$, the dot is empty and the reservoir energy distribution is Gaussian, with average $2k_{\rm B}T$ and width $0.1k_{\rm B}T$. The drive speeds are $\nu=\Gamma/100$ (blue lines) and $\Gamma/10$ (red lines).
}
\end{figure}

{\it Work extraction.}---To illustrate our approach, we first consider a basic protocol for work extraction. Starting at $t=0$ with an empty dot at energy $\epsilon_0$, we move the dot level down in energy with constant speed $\nu$ to zero, as $\epsilon_t=\epsilon_0(1-\nu t)$. By simulating a large number of trajectories, we obtain the statistical properties of the thermodynamic quantities. In Fig.~\ref{Fig2}\,a), the average extracted work $-W$, as well as the work extracted along individual trajectories, are shown as functions of time for different speeds. We see that $-W$, as well as the number of work extraction intervals, decrease for increasing $\nu$. 

The corresponding full probability distribution of extracted work is shown in Fig.~\ref{Fig2}\,b). For the fast drive, a sizeable fraction of trajectories display no electron tunneling and, hence, no work is extracted. For the slow drive the distribution becomes Gaussian shaped. Comparing to the work distribution of an infinite-size reservoir, the finite-size effects are most clearly visible for a slow drive, where they lead to a shift of the distribution towards smaller work values. 
Thus, the largest difference between the average work extracted with a finite and infinite-size reservoir seems to occur in the quasi-static regime.

In Fig.~\ref{Fig2}\,c), the distributions of the entropic temperature $T_\text e$ at the end of the protocol for the same parameters as in Fig.~\ref{Fig2}\,b) are shown. Compared to the distribution for large speed, the small speed distribution is narrowed and shifted to lower temperatures. In Fig.~\ref{Fig2}\,d) it is shown how the distribution of total entropy production is narrowed and shifted towards zero when the drive speed is decreased.

\begin{figure}
\includegraphics[width=2.4in]{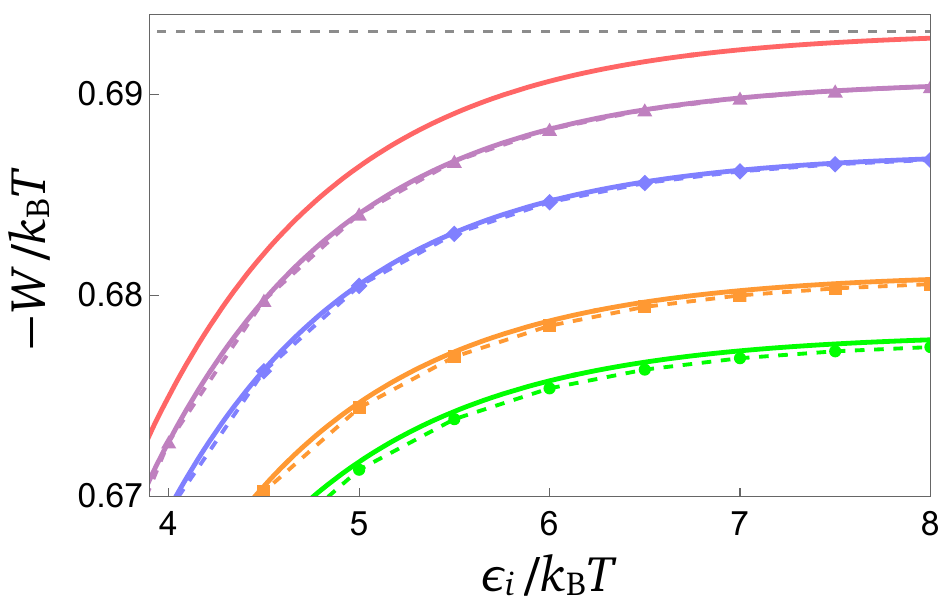}
\caption{\label{Fig3} 
Work extracted, $-W$, as a function of  $\epsilon_{\rm i}$ 
in a quasistatic cycle of the Szilard engine, for the following initial heat capacities $C(T(E_0))$: $16 k_\text B$ (green line), $20 k_\text B$ (orange line), $40 k_\text B$ (blue line), $100 k_\text B$ (purple line). The full numerical solution (solid lines) and the expansion to first order in $1/C(T(E_0))$ (dashed lines) are shown. The infinite-size reservoir case is shown with a solid, red line. { The dashed, gray line corresponds to the upper bound on work extraction with an infinite-size reservoir in a cycle of the Szilard engine, which is $k_\text B T\ln 2$.} Corresponding analytical expressions are given in the SI.
}
\end{figure}

To highlight the effect of a finite-size reservoir on information to work conversion, we analyze a Szilard engine, following closely the quantum dot protocol in Ref.~\cite{Barker2022}. Initially, the dot level energy is put to zero, giving a dot occupation probability 1/2, and the reservoir energy is fixed to $E_0$. The occupation is then measured, with two possible outcomes: i) If the dot is empty, the level energy is instantaneously increased to $\epsilon_{\rm i}$ and thereafter quasi-statically taken back to zero. ii) If the dot  is instead occupied by an electron, the level energy is  instantaneously decreased to $-\epsilon_{\rm i}$, and thereafter quasi-statically increased back to zero. We note that the process is not completely cyclic, since the initial reservoir energy is well-defined, while the final energy is a stochastic quantity. The average extracted work $-W$ as a function of $\epsilon_0$, for different heat capacities, is shown in Fig.~\ref{Fig3}. We see that decreasing the size of the reservoir leads to a monotonically decreasing $-W$. 
This is in line with previous results showing that Landauer’s upper bound on work extraction cannot be achieved with finite-size reservoirs~\cite{Reeb2014}.

{\it Conclusion.---} We provided a consistent thermodynamic description for a two-level system,
namely a quantum dot, coupled to a finite-size reservoir.
In our approach, the reservoir entropy along a given trajectory is determined by the entropic temperature, which therefore dictates the thermodynamics of the system and finite-size reservoir. Notably, we found that the entropic temperature is required to describe the thermodynamics of the system and reservoir as long as energy exchange occurs in quanta. When energy quantization is neglible, the entropic temperature reduces to the actual temperature, and therefore a connection with Clausius' second law is established.
We complete our analysis by defining work and heat, and by showing that the stochastic entropy production vanishes for each trajectory in the quasi-static limit. Our results are illustrated by a protocol for work extraction and for the Szilard engine.

Our results show how to describe the thermodynamics of a finite-size reservoir that can be described by a fluctuating temperature. While we focus on an electronic two-level system, our results can easily be generalized to other scenarios, e.g., a superconducting qubit coupled to an electro-magnetic environment~\cite{Spiecker2023, SpieckerPavlov2023} or an electron spin coupled to nuclear spins \cite{Jackson2022}. Of particular interest are systems, or reservoirs, which exhibit quantum coherence, such as double quantum dots \cite{prech2023} or squeezed bosonic reservoirs \cite{Manzano2018}. Furthermore, our approach can be adapted to include multiple reservoirs, allowing for transport scenarios.

Finally, we note that a microscopic derivation of our master equation in Eq.~\eqref{MasterEquation} would provide quantitative insight into the limitation of our approach and is left for future work. A starting point for such a derivation could be provided by the extended micro-canonical master equation \cite{Esposito2003,Breuer2006,Esposito2007,Breuer2007,RieraCampeny2021,RieraCampeny2022}.


{\it Acknowledgements.---} S.V.M. and P.S. acknowledge support from the Knut and Alice Wallenberg Foundation (Project No.~2016-0089). S.V.M. acknowledges funding from the European Commission via the Horizon Europe project ASPECTS (Grant Agreement No.~101080167), and P.S. acknowledges support from the Swedish Research Council (Grant No.~2018-03921). P.P.P. acknowledges funding from the Swiss National Science Foundation (Eccellenza Professorial Fellowship PCEFP2\_194268).

\bibliography{references.bib}

\newpage
\widetext
\begin{center}
	\textbf{\large Supplemental information} 
\end{center}
\setcounter{equation}{0}
\setcounter{figure}{0}
\setcounter{table}{0}
\setcounter{page}{1}
\makeatletter
\renewcommand{\theequation}{S\arabic{equation}}
\renewcommand{\thefigure}{S\arabic{figure}}

In this supplemental information, we provide derivations and motivations for different concepts introduced in the main text. In Sec.~\ref{app:rates}, we motivate the rate-equation used in the main text, and we discuss charging energies and chemical potentials. Section \ref{app:entropicT} derives the fluctuation theorem introduced in the main text and the corresponding entropy production featuring the entropic temperature. In Sec.~\ref{app:quasistatic}, we derive expressions for various quantities in the quasi-static regime, in Sec.~\ref{app:qswork}, we provide additional details on work extraction in the quasi-static regime, deriving a perturbative result for large heat capacities, and in Sec.~\ref{app:entropies}, we consider previously derived entropy productions which are expected to be positive for our scenario.

\section{Tunneling rates, charging effects, and chemical potential}
\label{app:rates}
Here we motivate the rates given in Eq.~\eqref{TunnelingRateOUT} in the main text and we briefly discuss charging effects and the effect of a chemical potential of the finite-size reservoir.  As a starting point we consider a single level quantum dot tunnel coupled to a metallic island with fixed chemical potential and temperature. Within the sequential tunneling approximation, accounting for charging effects, the tunnel rates in and out of the quantum dot with a level energy $\xi_t$, are given by (see, e.g., Ref.~\cite{Golovach2011} for a system with a quantum dot coupled to a metallic island)
\begin{equation}
\label{eq:fermisupp}
\Gamma_{\rm in}=\Gamma f(\xi_t-\Delta), \hspace{0.5 cm} \Gamma_{\rm out}=\Gamma[1- f(\xi_t-\Delta)],
\end{equation} 
where $f(\epsilon)$ is the Fermi function, c.f.~ Eq.~\eqref{fermi}, and $\Delta$ is the total change in energy between the charge state with one electron on the quantum dot and N electrons on the reservoir and the charge state with 0 electrons on the dot and N+1 electrons on the reservoir. This energy $\Delta$ arises from Coulomb interactions between the electrons on the reservoir and between electrons on the reservoir and the electron on the dot. We note that the charging energy can be absorbed into the dot energy $\epsilon_t=\xi_t-\Delta$ since it enters $\Gamma_{\rm in}$ and $\Gamma_{\rm out}$ in an identical manner. This is the case because an electron entering the metallic island needs to provide the charging energy to enter and an electron leaving the island takes this energy with it (and because we only consider two charge states).

To take into account the fluctuations in the finite-size reservoir due to the exchange of energy and particles, both temperature and chemical potential fluctuations should be considered. Temperature fluctuations are included as discussed in the main text (and following Ref.~\cite{vandenBerg2015}) by replacing the fixed temperature in the Fermi function by $T(E)$ [c.f.~Eq.~\eqref{Temperature}], where $E$ is the fluctuating energy of the reservoir.
This is motivated by a separation of two time-scales $\tau_{e-e}\ll\tau_{E}$~\cite{vandenBerg2015}. The fast time-scale, $\tau_{e-e}$ describes the time-scale over which the reservoir thermalizes to the temperature $T(E)$ due to electron-electron interactions. This occurs much faster than the time-scale $\tau_E$ which governs the change of energy $E$ due to tunneling between the system and the reservoir and is the time-scale which is described by the tunneling rates given in Eq.~\eqref{TunnelingRateOUT} in the main text.

To include the chemical potential fluctuations, we note that the chemical potential is by definition equal to the energy cost of adding one electron to the reservoir. This energy cost contains two contributions: the charging energy, as well as a contribution from the density of states at the Fermi energy $\mu_{\text 0}$. Here we treat these two contributions on an equal footing, which allows us to include the contribution from the density of states in the parameter $\Delta$ in Eq.~\eqref{eq:fermisupp}. Finally, we absorb $\Delta$ into the dot energy $\epsilon_t=\xi_t-\Delta$, which results in the tunneling rates given in Eq.~\eqref{TunnelingRateOUT} in the main text. Effects that go beyond treating the two contributions to the chemical potential on an equal footing provide a subject for interesting follow up studies, which go beyond the scope of this work.

\section{Entropic temperature from a fluctuation theorem}
\label{app:entropicT}
Here we present a derivation of the fluctuation theorem and the entropic temperature presented in the main text in Eqs.~\eqref{FluctTheorem} and \eqref{EffectiveT}. We follow similar steps as in previous works, such as Ref.~\cite{vandenBroeck2015}. For clarity, we start with a process with a single tunnel event. Consider a trajectory $\gamma$ in which, at $t=0$, the dot starts empty, $n_0 = 0$, and the reservoir energy is $E_0$. During the time interval $0\leq t<\tau_1$ there is no tunnel event. Then, at time $t = \tau_1$, an electron jumps from the reservoir into the dot. The electron then remains on the dot during the time interval $\tau_1<t\leq\tau$, until the end of the process at $t=\tau$. That is, at $t=\tau$ the final state of the system is $n_\tau = 1$, and the reservoir energy is $E_0 - \epsilon_{\tau_1}$. Given the probability $p(n_0, E_0;0)$ that the dot is empty  at $t=0$, the probability density $P(\gamma)$ of observing the trajectory $\gamma$ can then be expressed as
\begin{equation}\label{ProbForward}
P(\gamma) = p(n_0,E_0;0) e^{-\int_0^{\tau_1}  \Gamma_{\text{in}}(\epsilon_t,E_0) dt} \Gamma_{\text{in}}(\epsilon_{\tau_1},E_0) e^{-\int_{\tau_1}^{\tau}  \Gamma_{\text{out}}(\epsilon_t,E_0 - \epsilon_{\tau_1}) dt}.
\end{equation}
The probability $\tilde P(\tilde{\gamma})$ for the time-reversed trajectory $\tilde \gamma$ can, along the same lines, be written as
\begin{equation}\label{ProbBackward}
\tilde P(\tilde{\gamma}) = \tilde{p}(n_\tau, E_0 - \epsilon_{\tau_1};0) e^{-\int_0^{\tau - \tau_1} \tilde \Gamma_{\text{out}}(\epsilon_{t'},E_0 - \epsilon_{\tau-\tau_1}) dt^\prime} \tilde \Gamma_{\text{out}}(\epsilon_{\tau - \tau_1},E_0 - \epsilon_{\tau - \tau_1}) e^{-\int_{\tau - \tau_1}^{\tau}  \tilde \Gamma_{\text{in}}(\epsilon_{t^{\prime}},E_0) dt^\prime}.
\end{equation}
where $\tilde{p}(n_\tau, E_0 - \epsilon_{\tau_1};0)$ is the probability that the dot is occupied by an electron at the start of the time-reversed trajectory. The $\tilde ~$ on the tunnel rates denotes that the time evolution of the dot-level should be reversed when evaluating the rates. Now, since the times along the forward and backward trajectories are related as $t^\prime = \tau - t$, the probability in Eq.~\eqref{ProbBackward} can be rewritten as
\begin{equation}\label{ProbBackward2}
\tilde P(\tilde{\gamma}) = p(n_\tau, E_0 - \epsilon_{\tau_1};\tau) e^{-\int_{\tau_1} ^{\tau}  \Gamma_{\text{in}}(\epsilon_t,E_0 - \epsilon_{\tau_1}) dt} \Gamma_{\text{out}}(\epsilon_{\tau_1},E_0 - \epsilon_{\tau_1}) e^{-\int_{0}^{\tau}  \Gamma_{\text{out}}(\epsilon_t,E_0 ) dt},
\end{equation}
where we chose the initial probability of the backward experiment to equal the final probability of the forward experiment, $\tilde{p}(n_\tau, E_0 - \epsilon_{\tau_1};0)=p(n_\tau, E_0 - \epsilon_{\tau_1};\tau)$.
We can thus write the ratio of the forward and backward probabilities as
\begin{equation}\label{1Jump}
    \frac{P(\gamma)}{\tilde P(\tilde{\gamma})} = \frac{p(n_0,E_0;0)}{p(n_\tau,E_0 - \epsilon_{\tau_1};\tau)}\frac{\Gamma_{\text{in}}(\epsilon_{\tau_1},E_0)}{\Gamma_{\text{out}}(\epsilon_{\tau_1},E_0 - \epsilon_{\tau_1})}.
\end{equation}
noting that the terms describing the probabilities for no tunnel event to occur are the same for both trajectories and hence drop out of the ratio.  

This result can now be extended to an arbitrary trajectory $\gamma$ with $M$ jumps at times $0\leq \tau_1 \leq \tau_2.....\tau_{M-1}\leq \tau_M$. The trajectory starts at time $t=0$ in state $n_0$ at energy $E_0$ and ending at time $t=\tau$, in $n_\tau$ at $E_\tau$. The ratio of the forward and backward probabilities can then directly be written (for $M$ odd)
\begin{equation}\label{MJumps}
\frac{P(\gamma)}{\tilde P(\tilde{\gamma})} = \frac{p(n_0,E_0;0)}{p(n_\tau,E_\tau;\tau)}\frac{\Gamma_{\text{in}}(\epsilon_{\tau_1},E_0)}{\Gamma_{\text{out}}(\epsilon_{\tau_1},E_0 - \epsilon_{\tau_1})}\frac{\Gamma_{\text{out}}(\epsilon_{\tau_2},E_0-\epsilon_{\tau_1})}{\Gamma_{\text{in}}(\epsilon_{\tau_2},E_0 - \epsilon_{\tau_1}+\epsilon_{\tau_2})}...\frac{\Gamma_{\text{in}}(\epsilon_{\tau_M},E_0-\epsilon_{\tau_1}+....+\epsilon_{\tau_{M-1}},\tau_M)}{\Gamma_{\text{out}}(\epsilon_{\tau_M},E_0 - \epsilon_{\tau_1}+....-\epsilon_{\tau_M})}.
\end{equation}
To connect to Eq.~(\ref{FluctTheorem}) in the main text, this ratio can be written as
\begin{equation}
\frac{P(\gamma)}{\tilde P(\tilde{\gamma})} =\exp\left[\frac{\Delta s+\Delta s_{\rm r}}{k_\text B}\right],
\end{equation}
where
\begin{equation}
\Delta s=k_\text B \left[\ln p(n_0,E_0;0)-\ln p(n_\tau,E_\tau;\tau)\right],
\end{equation}
the expression for the change in the system entropy, Eq.~(\ref{EntrSystem}) in the main text, and
\begin{equation}
\Delta s_{\rm r}=k_\text B\left(-\ln\left[\frac{\Gamma_{\text{out}}(\epsilon_{\tau_1},E_0 - \epsilon_{\tau_1},\tau_1)}{\Gamma_{\text{in}}(\epsilon_{\tau_1},E_0)}\right]+\ln\left[\frac{\Gamma_{\text{out}}(\epsilon_{\tau_2},E_0-\epsilon_{\tau_1})}{\Gamma_{\text{in}}(\epsilon_{\tau_2},E_0 - \epsilon_{\tau_1}+\epsilon_{\tau_2})}\right]....- \ln\left[ \frac{\Gamma_{\text{out}}(\epsilon_{\tau_M},E_0 - \epsilon_{\tau_1}+....-\epsilon_{\tau_M})}  {\Gamma_{\text{in}}(\epsilon_{\tau_M},E_0-\epsilon_{\tau_1}+....+\epsilon_{\tau_{M-1}})}\right]  \right).                  
\end{equation}
We then define the entropic temperature $T_\text e(t)$ such that
\begin{equation}\label{sr}
\Delta s_{\rm r}=k_\text B\left( \sum_{j=1}^{M}\frac{(-1)^j\epsilon_{\tau_j}}{k_{\rm B} T_\text e(\tau_j)} \right)
\end{equation}
that is,
\begin{equation}
T_\text e(\tau_1)=\frac{\epsilon_{\tau_1}}{k_\text B}\left(\ln\left[\frac{\Gamma_{\text{out}}(\epsilon_{\tau_1},E_0 - \epsilon_{\tau_1})}{\Gamma_{\text{in}}(\epsilon_{\tau_1},E_0)}\right]\right)^{-1}, \hspace{0.5cm} T_\text e(\tau_2)=\frac{\epsilon_{\tau_2}}{k_\text B}\left(\ln\left[\frac{\Gamma_{\text{out}}(\epsilon_{\tau_2},E_0-\epsilon_{\tau_1})}{\Gamma_{\text{in}}(\epsilon_{\tau_2},E_0 - \epsilon_{\tau_1}+\epsilon_{\tau_2})}    \right]\right)^{-1},...
\end{equation}
and similar for the other $T_\text e(\tau_j)$. As a final step we extend the definition $T_\text e(t)$ to be valid for all times, and not only the times when an electron tunnels, by writing it in terms of the total energy $\mathcal{E}(t)=E(t)+n(t)\epsilon_t$ along the trajectory, as
\begin{equation}
T_\text e(t)=\frac{\epsilon_{t}}{k_\text B}\left(\ln\left[\frac{\Gamma_{\text{out}}(\epsilon_t,\mathcal{E}(t) - \epsilon_t)}{\Gamma_{\text{in}}(\epsilon_t,\mathcal{E}(t))}\right]\right)^{-1},
\label{Te}
\end{equation}
which is the expression in Eq.~(\ref{EffectiveT}) of the main text. With this definition we can write the entropy $\Delta s_{\rm r}$ in Eq. (\ref{sr}) as the stochastic integral in Eq.~(\ref{EntrReservoir}) in the main text.

\section{Thermodynamic quantities in the quasi-static limit}
\label{app:quasistatic}
Here we present the derivation of the different thermodynamical quantities in the quasi-static limit. For a sufficiently slow drive of the dot level energy, we find that the total energy $\mathcal{E}(t)=E(t)+\epsilon_t n(t)$ becomes a deterministic quantity and the dot occupation is approximately always in a thermal state at the corresponding entropic temperature given by $p_1^{\text s}(\epsilon_t|\mathcal{E}(t))$, c.f., Eq.~\eqref{Stationary}. We start by introducing a time-scale $\theta$ that is long compared to the tunneling rates but short compared to the rate of change of the entropic temperature, i.e.,
\begin{equation}
    \label{eq:thetaquasistat}
    \theta \Gamma_{\rm in/out}(\epsilon_t,E(t))\gg 1,\hspace{2cm}p_1^{\text s}(\epsilon_{t+\theta}|\mathcal{E}(t+\theta))\simeq p_1^{\text s}(\epsilon_{t}|\mathcal{E}(t)).
\end{equation}
These inequalities ensure that the occupation along a single trajectory is well approximated by a thermal state for the given entropic temperature at all times.
We may then approximate integrals of the form
%
%
%
\begin{equation}
\label{eq:intquasistat}
   \int_{t}^{t+\theta}dt' n(t')f(t') \simeq \theta p_1^{\text s}(\epsilon_{t}|\mathcal{E}(t))f(t) , \hspace{.5cm}\Rightarrow\hspace{.5cm}  \int_{t_1}^{t_2}dt n(t)f(t) \simeq \int_{t_1}^{t_2}dt p_1^{\text s}(\epsilon_{t}|\mathcal{E}(t))f(t),
\end{equation}
where $f(t)$ is any function that varies slowly on the time-scale $\theta$ and the second identity is obtained by splitting the integral into integrals over the time $\theta$ and then letting $\theta\rightarrow dt$. Thus, an average over time $\theta$ is equivalent to an ensemble average, for a fixed entropic temperature. For the work, in Eq. (\ref{StochWorkHeat}), the integral can thus be written as
%
%
%
\begin{equation}
w=\int_0^{\tau} \dot \epsilon_t n(t)dt=
\int_0^{\tau}\dot{\epsilon}_{t} p_1^{\text s}(\epsilon_{t}|\mathcal{E}(t))dt.
\label{work2}
\end{equation}
For the heat, in Eq. (\ref{StochWorkHeat}), the stochastic integral can be written using partial integration
\begin{equation}
q=\int_{\gamma}\epsilon_t dn(t)=n(\tau)\epsilon_{\tau}-n(0)\epsilon_0-\int_0^{\tau}n(t) \dot \epsilon_t dt=n(\tau)\epsilon_{\tau}-n(0)\epsilon_0-\int_0^{\tau}\dot{\epsilon}_tp_1^{\text s}(\epsilon_{t}|\mathcal{E}(t))  dt,
\label{heat}
\end{equation}
where we used $d[n(t)\epsilon_t]=\epsilon_t dn(t)+n(t) d\epsilon_t$ and $d\epsilon_t=\dot{\epsilon}_t dt$.
We thus recover the first law of thermodynamics
\begin{equation}
q+w=\Delta u, \hspace{0.5cm} \Delta u=n(\tau)\epsilon_{\tau}-n(0)\epsilon_0,
\end{equation}
where $\Delta u$ is the difference between the final and initial internal energy of the system.

We note that the energy of the reservoir $E(t)$ is given by the initial energy $E(0)$ minus the heat $q(t)$ extracted until time $t$, i.e. $E(t)=E(0)-q(t)$. In the quasi-static limit, from Eq.~\eqref{heat}, this becomes
\begin{equation}
E(t)=E(0)-n(t)\epsilon_{t}+n(0)\epsilon_0+\int_0^{t} \dot{\epsilon}_{t'}p_1^{\text s}(\epsilon_{t'}|\mathcal{E}(t')) dt'
\end{equation}

These quantities all depend on the total energy $\mathcal{E}(t)$. In the quasi-static regime, it can be written as
\begin{equation}
\label{eq:totedet}
\mathcal{E}(t)=E(t)+n(t)\epsilon_t=E(0)+n(0)\epsilon_0+\int_0^{t}\dot{\epsilon}_{t'} p_1^{\text s}(\epsilon_{t'}|\mathcal{E}(t'))dt'=\mathcal{E}(0)+\int_0^{t}\dot{\epsilon}_{t'} p_1^{\text s}(\epsilon_{t'}|\mathcal{E}(t'))dt'.
\end{equation}
This implies that for a given initial total energy $\mathcal{E}(0)$, the total energy $\mathcal{E}(t)$ becomes a deterministic quantity in the quasi-static regime. Furthermore, the entropic temperature in Eq.~\eqref{EffectiveT}, which is a function of $\mathcal{E}(t)$ and $\epsilon_t$, also becomes deterministic along a single trajectory. If we start with a distribution of initial total energies, work and total energy may vary from trajectory to trajectory, but they are completely determined by the initial value of the total energy.

We now turn to the evaluation of entropies in the quasi-static regime. We start with the change in system entropy
\begin{equation}
\label{eq:appentcangesys}
    \Delta s = - k_{\rm B}\ln p(n(\tau),E(\tau);\tau)+k_{\rm B}\ln p(n(0),E(0);0),
\end{equation}
where $p(n,E;t)$ is the solution to the master equation in Eq.~\eqref{MasterEquation}. In the quasi-static regime, these probabilities are well approximated by
\begin{equation}
\label{eq:probqs}
    p(n(t),E(t);t) = p_{n(t)}^{\text s}(\epsilon_{t}|\mathcal{E}(t)) Q_t(\mathcal{E}(t)),
\end{equation}
where $Q_t(\mathcal{E}(t))$ denotes the marginal probability for the total energy at time $t$. Since total energy is deterministic along a single trajectory, $Q_t(\mathcal{E})$ is fully determined by the initial distribution of total energies
\begin{equation}
    \label{eq:totalenergydist}
    Q_t(\mathcal{E}(t))d\mathcal{E}(t) = Q_0(\mathcal{E}(0))d\mathcal{E}(0).
\end{equation}
Importantly, the differential $d\mathcal{E}(t)$ changes over time. Thus, even if $\mathcal{E}(t)$ behaves deterministic, the distribution $Q_t(\mathcal{E}(t))$ may still broaden over time, contributing to the entropy change in Eq.~\eqref{eq:appentcangesys}. From Eq.~\eqref{eq:totedet}, we find
\begin{equation}
    \label{eq:totender}
    \frac{d\mathcal{E}(t)}{d\mathcal{E}(0)} = 1+\int_0^{t}\frac{d\mathcal{E}(t')}{d\mathcal{E}(0)}\dot{\epsilon}_{t'} \frac{dp_1^{\text s}(\epsilon_{t'}|\mathcal{E}(t'))}{d\mathcal{E}(t')}dt'.
\end{equation}
Taking the time-derivative of this equation, we find
\begin{equation}
    \frac{d}{dt} \frac{d\mathcal{E}(t)}{d\mathcal{E}(0)} =\dot{\epsilon}_{t} \frac{dp_1^{\text s}(\epsilon_{t}|\mathcal{E}(t))}{d\mathcal{E}(t)}\frac{d\mathcal{E}(t)}{d\mathcal{E}(0)}
    \hspace{.5cm}\Leftrightarrow\hspace{.5cm}\frac{d\mathcal{E}(t)}{d\mathcal{E}(0)} = e^{\int_0^t dt' \dot{\epsilon}_{t'} \frac{dp_1^{\text s}(\epsilon_{t'}|\mathcal{E}(t'))}{d\mathcal{E}(t')}dt'}.
\end{equation}
We thus find
\begin{equation}
    \label{eq:totalentime}
    Q_t(\mathcal{E}(t)) = Q_0(\mathcal{E}(0))e^{-\int_0^t dt' \dot{\epsilon}_{t'} \frac{dp_1^{\text s}(\epsilon_{t'}|\mathcal{E}(t'))}{d\mathcal{E}(t')}dt'}.
\end{equation}
Using this relation, together with Eq.~\eqref{eq:probqs}, in Eq.~\eqref{eq:appentcangesys}, we find
\begin{equation}
    \label{eq:appsysent}
    \begin{aligned}
    \Delta s &= -k_{\rm B}\ln \frac{p_{n(\tau)}^{\text s}(\epsilon_{\tau}|\mathcal{E}(\tau))}{p_{n(0)}^{\text s}(\epsilon_{0}|\mathcal{E}(0))} +k_{\rm B}\int_0^\tau dt \dot{\epsilon}_{t} \frac{dp_1^{\text s}(\epsilon_{t}|\mathcal{E}(t))}{d\mathcal{E}(t)}dt\\&= -k_{\rm B}\ln \frac{p_{n(\tau)}^{\text s}(\epsilon_{\tau}|\mathcal{E}(\tau))}{p_{n(0)}^{\text s}(\epsilon_{0}|\mathcal{E}(0))} +\int_0^\tau dt \dot{\epsilon}_{t}\epsilon_{t}p_1^{\text s}(\epsilon_{t}|\mathcal{E}(t))p_0^{\text s}(\epsilon_{t}|\mathcal{E}(t))\left\{\frac{1-f(\epsilon_{t},\mathcal{E}(t))}{2T(\mathcal{E}(t))\mathcal{E}(t)}+\frac{f(\epsilon_{t},\mathcal{E}(t)-\epsilon_{t})}{2T(\mathcal{E}(t)-\epsilon_{t})[\mathcal{E}(t)-\epsilon_{t}]}\right\}.
    \end{aligned}
\end{equation}

\begin{figure}
\includegraphics[width=.8\textwidth]{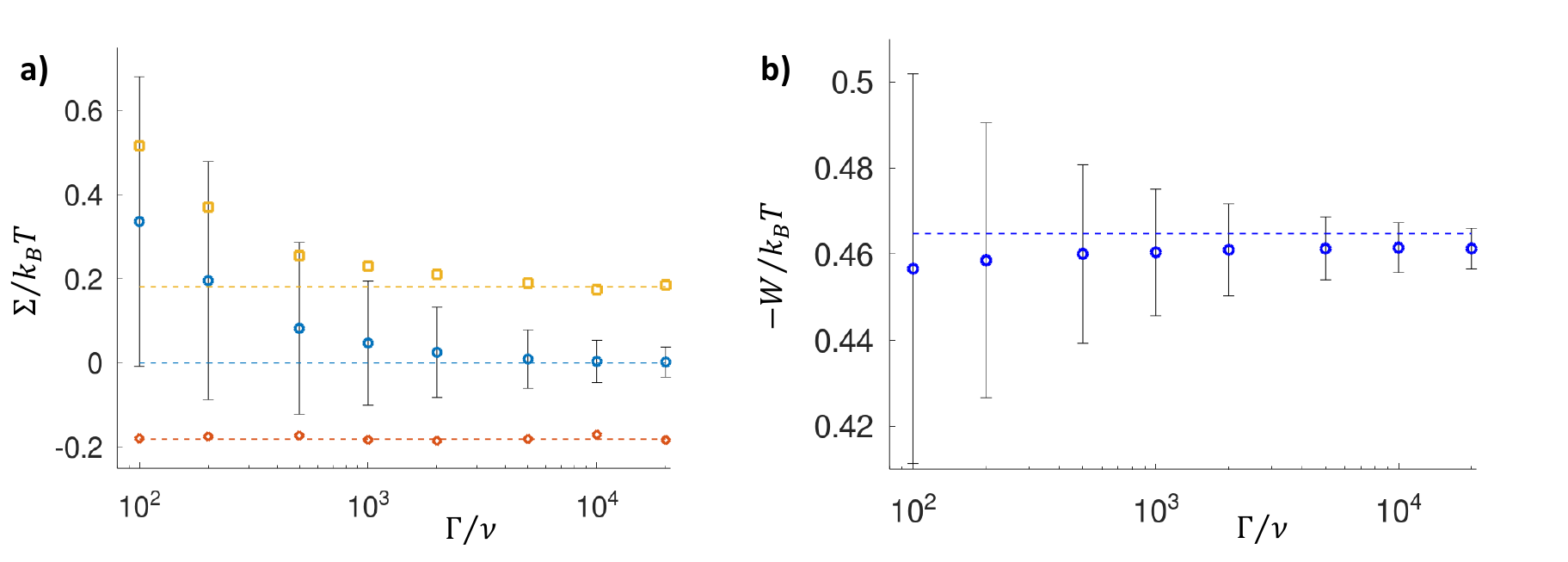}
\caption{Protocol speed dependence of entropy and work. The dot level is driven as $\epsilon_t=\epsilon_0(1-\nu t)$ with constant speed $\nu$. Initially, at $t=0$, the dot and the reservoir are uncorrelated, and the dot is populated with a probability $p_1^s(\epsilon_0|2k_{\rm B}T)$. The initial reservoir energy distribution is Gaussian with a mean energy $2k_{\rm B}T$ and a width $0.1 k_{\rm B}T$, with $T$ the initial reservoir temperature (for the mean energy). The initial heat capacity is $C=4k_{\rm B}$ and the initial dot energy $\epsilon_0=1.5k_{\rm B}T$. a) Average total entropy $\Sigma=\langle \sigma \rangle=\langle \Delta s+\Delta s_{\rm r}\rangle$ is shown (blue circles) for a set of protocol run times, or inverse protocol speeds, $\Gamma/\nu =100,200,500,1000,2000,5000,10000,20000$. The standard deviation $\sqrt{\langle (\sigma -\langle \sigma\rangle)^2\rangle}$ is shown as error bars. As a reference, the value of the average total entropy in the quasi-static limit, $\Sigma=0$, is shown as a blue, dashed line. The average entropy change of the reservoir $\langle \Delta s_{\rm r}\rangle$ (red diamonds) and the difference in system entropy $\Delta S=\langle \Delta s\rangle$ (yellow squares) are shown for the same protocol speeds, with their quasi-static limit values denoted by red and yellow dashed lines, respectively. b) Average extracted work $W=\langle w \rangle $ (blue circles) for the same protocol speeds as in a), with the standard deviation $\sqrt{\langle (w -\langle w\rangle)^2\rangle}$ shown by error bars. As a reference, the average work in the quasi-static limit is shown as a dashed blue line. } \label{fig:si}
\end{figure}

For the reservoir entropy production $\Delta s_\text r$, we can write the stochastic integral in Eq.~(\ref{EntrReservoir}) as
\begin{equation}
\Delta s_\text r=-\int_\gamma \frac{dq(t)}{T_\text e(t)}=-\int_\gamma \frac{\epsilon_tdn(t)}{T_\text e(t)}=- \frac{n(\tau)\epsilon_{\tau}}{T_\text e(\tau)}+\frac{n(0)\epsilon_{0}}{T_\text e(0)}+\int_0^{\tau} n(t)d\left[\frac{\epsilon_t}{T_\text e(t)}\right],
\end{equation}
performing a partial integration in the last step. We note that the last integral is \textit{not} of the form of Eq.~\eqref{eq:intquasistat} because the increment multiplying $n(t)$ depends itself on the value of $n(t)$. To make progress, we note that the entropic temperature does not change during a jump in $n(t)$. Furthermore, when $n(t)=1$, the total energy changes by $\dot{\epsilon}_t dt$ in the time-interval $dt$. This allows us to write
\begin{equation}
    \label{eq:discincte}
    \begin{aligned}
    n(t)d\left[\frac{\epsilon_t}{T_\text e(t)}\right] =n(t)k_\text B \dot{\epsilon}_t dt\left\{\left(\frac{\partial\ln\frac{p_0^{\text s}(\epsilon_t|\mathcal{E}(t))}{p_1^{\text s}(\epsilon_t|\mathcal{E}(t))}}{\partial\epsilon_t}\right)_{\mathcal{E}(t)}+\left(\frac{\partial\ln\frac{p_0^{\text s}(\epsilon_t|\mathcal{E}(t))}{p_1^{\text s}(\epsilon_t|\mathcal{E}(t))}}{\partial\mathcal{E}(t)}\right)_{\epsilon_t}\right\},
    \end{aligned}
\end{equation}
where the subscripts imply that we keep $\mathcal{E}(t)$ constant when taking the derivative with respect to $\epsilon_t$ and vice versa and we used
\begin{equation}
\label{eq:teapp}
T_\text e(t)=\frac{\epsilon_{t}}{k_\text B}\left(\ln\left[\frac{p_0^\text s(\epsilon_t|\mathcal{E}(t))}{p_1^\text s(\epsilon_t|\mathcal{E}(t))}\right]\right)^{-1}.
\end{equation}
We may now use Eq.~\eqref{eq:intquasistat} to obtain
\begin{equation}
    \label{eq:intcompl}
    \int_0^{\tau} n(t)d\left[\frac{\epsilon_t}{T_\text e(t)}\right] = k_\text B\int_0^\tau dtp_1^{\text s}(\epsilon_t|\mathcal{E}(t)) \dot{\epsilon}_t \left\{\left(\frac{\partial\ln\frac{p_0^{\text s}(\epsilon_t|\mathcal{E}(t))}{p_1^{\text s}(\epsilon_t|\mathcal{E}(t))}}{\partial\epsilon_t}\right)_{\mathcal{E}(t)}+\left(\frac{\partial\ln\frac{p_0^{\text s}(\epsilon_t|\mathcal{E}(t))}{p_1^{\text s}(\epsilon_t|\mathcal{E}(t))}}{\partial\mathcal{E}(t)}\right)_{\epsilon_t}\right\}.
\end{equation}
Using the quasi-static expression for the total energy in Eq.~\eqref{eq:totedet}, we find $\dot{\mathcal{E}}(t) = p_1^{\text s}(\epsilon_t|\mathcal{E}(t))\dot{\epsilon}_t$. Together with the identity
\begin{equation}
    \frac{d}{dt}\ln\frac{p_0^{\text s}(\epsilon_t|\mathcal{E}(t))}{p_1^{\text s}(\epsilon_t|\mathcal{E}(t))} = \dot{\epsilon}_t\left(\frac{\partial\ln\frac{p_0^{\text s}(\epsilon_t|\mathcal{E}(t))}{p_1^{\text s}(\epsilon_t|\mathcal{E}(t))}}{\partial\epsilon_t}\right)_{\mathcal{E}(t)}+\dot{\mathcal{E}}(t)\left(\frac{\partial\ln\frac{p_0^{\text s}(\epsilon_t|\mathcal{E}(t))}{p_1^{\text s}(\epsilon_t|\mathcal{E}(t))}}{\partial\mathcal{E}(t)}\right)_{\epsilon_t},
\end{equation}
this results in
\begin{equation}
    \label{eq:intcompl2}
    \int_0^{\tau} n(t)d\left[\frac{\epsilon_t}{T_\text e(t)}\right] = k_\text B\int_0^\tau dtp_1^{\text s}(\epsilon_t|\mathcal{E}(t))  \frac{d}{dt}\ln\frac{p_0^{\text s}(\epsilon_t|\mathcal{E}(t))}{p_1^{\text s}(\epsilon_t|\mathcal{E}(t))}+k_\text B\int_0^\tau dtp_1^{\text s}(\epsilon_t|\mathcal{E}(t))p_0^{\text s}(\epsilon_t|\mathcal{E}(t))\dot{\epsilon}_t\left(\frac{\partial\ln\frac{p_0^{\text s}(\epsilon_t|\mathcal{E}(t))}{p_1^{\text s}(\epsilon_t|\mathcal{E}(t))}}{\partial\mathcal{E}(t)}\right)_{\epsilon_t}.
\end{equation}
We may now show that $\Delta s_{\rm r} = -\Delta s$ by using
\begin{equation}
    \label{eq:deltsid1}
    - \frac{n(\tau)\epsilon_{\tau}}{T_\text e(\tau)}+\frac{n(0)\epsilon_{0}}{T_\text e(0)} + k_\text B\int_0^\tau dtp_1^{\text s}(\epsilon_t|\mathcal{E}(t))  \frac{d}{dt}\ln\frac{p_0^{\text s}(\epsilon_t|\mathcal{E}(t))}{p_1^{\text s}(\epsilon_t|\mathcal{E}(t))} = k_{\rm B}\ln \frac{p_{n(\tau)}^{\text s}(\epsilon_{\tau}|\mathcal{E}(\tau))}{p_{n(0)}^{\text s}(\epsilon_{0}|\mathcal{E}(0))},
\end{equation}
where we used Eq.~\eqref{eq:teapp}, together with
\begin{equation}
    \label{eq:deltsid2}
    k_\text B\int_0^\tau dtp_1^{\text s}(\epsilon_t|\mathcal{E}(t))p_0^{\text s}(\epsilon_t|\mathcal{E}(t))\dot{\epsilon}_t\left(\frac{\partial\ln\frac{p_0^{\text s}(\epsilon_t|\mathcal{E}(t))}{p_1^{\text s}(\epsilon_t|\mathcal{E}(t))}}{\partial\mathcal{E}(t)}\right)_{\epsilon_t}=-k_{\rm B}\int_0^\tau dt \dot{\epsilon}_{t} \frac{dp_1^{\text s}(\epsilon_{t}|\mathcal{E}(t))}{d\mathcal{E}(t)}dt.
\end{equation}
Comparing the sum of the right-hand sides of Eqs.~\eqref{eq:deltsid1} and \eqref{eq:deltsid2} to the first line of Eq.~\eqref{eq:appsysent}, we find $\Delta s_{\rm r} = -\Delta s$, implying that in the quasi-static regime, the total entropy production vanishes for each trajectory. 
Figure \ref{fig:si} illustrates the average work and total entropy production together with their variances, as well as the averages of the system and reservoir contribution to the entropy production for different protocol speeds. The numerical results agree well with the analytical results derived in this section.

\section{Quasi-static work extraction for large heat capacities}
\label{app:qswork}
Here we present a derivation of the average work $W$ during a time interval $0 \leq t\leq \tau$, focusing on the quasi-static limit. From Eq. (\ref{work2}) we have  
\begin{equation}\label{WorkExtracted}
 W = \int_0^\tau  \dot{\epsilon_t} p_1^{\text s}(\epsilon_t|\mathcal{E}(t)) dt.
 \end{equation}
Changing the integration variable in Eq. (\ref{WorkExtracted}), using that $\dot{\epsilon_t}dt = d\epsilon$, we have
\begin{equation}\label{WorkExtractedI}
 W = \int_{\epsilon_0}^{\epsilon_{\tau}}  p_1^{\text s}(\epsilon|\mathcal{E}(\epsilon)) d\epsilon =   \int_{\epsilon_0}^{\epsilon_\tau}  \frac{f(\epsilon, \mathcal{E}(\epsilon))}{1 - f(\epsilon, \mathcal{E}(\epsilon)-\epsilon) + f(\epsilon, \mathcal{E}(\epsilon))} d\epsilon,
 \end{equation}
where we expressed the total energy as a function of $\epsilon$
\begin{equation}
    \label{eq:toteeps}
    \mathcal{E}(\epsilon) =\mathcal{E}(\epsilon_0)+W(\epsilon_0,\epsilon),\hspace{1cm}W(\epsilon_0,\epsilon)=\int_{\epsilon_0}^{\epsilon}  \frac{f(\epsilon', \mathcal{E}(\epsilon'))}{1 - f(\epsilon', \mathcal{E}(\epsilon')-\epsilon') + f(\epsilon', \mathcal{E}(\epsilon'))} d\epsilon'.
\end{equation}
%
The full integral in Eq.~(\ref{WorkExtracted}) is evaluated numerically to obtain the curves in Fig.~\ref{Fig2} in the main text. 

In the limit of large heat capacity, when the temperature fluctuations are small, it is possible to arrive at an analytical expression for the extracted work. To this end we first introduce dimensionless quantities as $\tilde{C}_T 
\equiv C(T)/k_{\rm B}, \tilde{W} \equiv W/k_{\rm B} T, \tilde{\epsilon} \equiv \epsilon/k_{\rm B} T, \tilde \epsilon_0=\epsilon_0/k_{\rm B}T,\tilde \epsilon_\tau=\epsilon_\tau/k_{\rm B}T$, where $T\equiv T(E_0)$ denotes the initial temperature of the reservoir 
%
Expanding the integrand in Eq. (\ref{WorkExtractedI}) to first order in $1/\tilde C_{T}$ gives
\begin{equation}\label{Approx}
\frac{f(\epsilon, \mathcal{E}(\epsilon))}{1 - f(\epsilon, \mathcal{E}(\epsilon)-\epsilon) + f(\epsilon, \mathcal{E}(\epsilon))}  \approx  f(\tilde \epsilon)-\frac{f(\tilde \epsilon) [1-f(\tilde \epsilon)]\tilde \epsilon}{\tilde C_T}\left[f(\tilde \epsilon)\tilde{\epsilon}- \tilde{W}_{\rm inf}(\tilde \epsilon_0,\tilde \epsilon)\right]
\end{equation}
where we abbreviated $f(\tilde \epsilon)=1/[1+\exp(\tilde \epsilon)]$ and introduced the (dimensionless) work corresponding to an infinite reservoir as
\begin{equation}\label{Winf}
 \tilde{W}_{\text{inf}}(\tilde \epsilon_0,\tilde \epsilon)= \ln\frac{1-f(\tilde \epsilon)}{1-f(\tilde \epsilon_0)} = \ln \frac{1+e^{-\tilde{\epsilon}_0}}{ 1 + e^{-\tilde{\epsilon}}}. 
\end{equation}
We can now perform the integral in Eq. (\ref{WorkExtractedI}) giving the work
\begin{equation}
\label{eq:extractedworkpert}
\tilde{W} \approx \tilde{W}_{\text{inf}} (\tilde \epsilon_0,\tilde \epsilon_\tau) - \frac{1}{2\tilde{C}_T}\left\{\left[\tilde{\epsilon}_\tau f(\tilde{\epsilon}_\tau)-\tilde{W}_{\text{inf}} (\tilde \epsilon_0,\tilde \epsilon_\tau)\right]^2-\tilde{\epsilon}_0^2f^2(\tilde{\epsilon}_0)\right\}.
\end{equation}

%

\section{Comparison of entropy expressions}
\label{app:entropies}
There are several expression for the entropy production for systems coupled to finite size reservoirs in the existing literature. Of particular relevance for our analysis are the works by Esposito, Lindenberg, and Van den Broeck~\cite{Esposito2010} as well as Strasberg, Díaz, and Riera-Campeny~\cite{Strasberg2021-2}, where the expressions for the total entropy production at time $t$,  $\Sigma_1(t)$ and $\Sigma_2(t)$ respectively, for a process starting at $t=0$, are given by (for a single reservoir)
\begin{equation}
\Sigma_1(t)= \Delta S_{\rm S}(t) - \frac{ Q(t)}{T}, \hspace{0.5cm} \Sigma_2(t)= \Delta S_{\rm S}(t) - \int_0^t dt' \frac{ \dot Q(t')}{T[Q(t')]}= \Delta S_{\rm S}(t)-C'\left(T[ Q(t)]-T\right).
\label{totentr}
\end{equation}  
Here $T[Q(t)]=2[(E(0)-Q(t))/C']^{1/2}$ with $Q(t)$ the time dependent, ensemble averaged heat, $E(0)$ the initial reservoir energy, and $T\equiv T[Q(0)]$ the initial reservoir temperature. Furthermore, $\Delta S_{\rm S}(t)$ is the change in entropy in the reduced system
\begin{equation}
\Delta S_{\rm S}(t)/k_B=p_1(0)\ln\left[p_1(0)\right]+[1-p_1(0)]\ln[1-p_1(0)]-\left\{p_1(t)\ln\left[p_1(t)\right]+[1-p_1(t)]\ln[1-p_1(t)]\right\},
\end{equation}  
where $p_1(t) = \int dE p(1,E;t)$ is the probability of the dot being occupied. We stress that $\Delta S_{\rm S}$ is not the ensemble average of the change in system entropy $\Delta s$ defined in Eq.~\eqref{EntrSystem}, which also takes into account the distribution of reservoir energies.
In Ref.~\cite{Strasberg2021-2} it was shown that
\begin{equation}
\label{eq:enthyr}
\Sigma_1(t) \geq \Sigma_2(t),
\end{equation} 
for any system and time $t$. We stress that the expressions in Eq. (\ref{totentr}) are derived under the sole assumptions that the initial state of the system and reservoir is a product state and that the reservoir starts out in a thermal state with temperature $T$.

To achieve a quantitative comparison between our expression for the total entropy, $\Sigma(t)$ in Eq. (\ref{EnsembleEntropyProduction}) in the main text, and $\Sigma_1(t)$ and $\Sigma_2(t)$, we consider a process where the dot level energy $\epsilon_t$ is quasistatically driven as a function of time $t$, starting at $t=0$. In this quasistatic limit $\Sigma(t)=0$. Moreover, we consider a large reservoir, with a heat capacity obeying $C_T/k_BT\gg 1$, allowing us to treat the effects of the finite reservoir size perturbatively, similar to the preceeding section. Under these conditions, the dot occupation is approximately $p_1(t)\simeq p_1^s(\epsilon_t \vert \mathcal{E}(\epsilon_t))$. To obtain expressions for the reservoir entropy production in $\Sigma_1(t)$ and $\Sigma_2(t)$ we first express the ensemble average heat, via the first law in Eq. (\ref{1stLaw}), as
\begin{equation}
Q(t)=-W(t)+U(t)-U(0),
\end{equation}  
where the performed work is $W(t)=\int_0^tdt' p_1(t')\dot \epsilon_{t'}$ and the system internal energy is $U(t)=p_1(t) \epsilon_t$. In the limit of large reservoir heat capacity at $t=0$, where $\tilde C_T=C_T/k_B\gg 1$, with $C_T=C'T$, we can expand all quantities to first order in $1/\tilde C_T$. For $p_1(t)$ and $W(t)$ this was done in the preceeding sections, in Eqs. (\ref{Approx}) and (\ref{eq:extractedworkpert}). To simplify the analysis below we introduce the following notation for the expanded probability, as
\begin{equation}
p_1(t)=f(\tilde \epsilon_t)+\frac{1}{\tilde C_T}\delta p_1(t), \hspace{0.5 cm} \hspace{0.5 cm} \delta p_1(t)=-f(\tilde \epsilon_t)[1-f(\tilde \epsilon_t)]\left(f(\tilde \epsilon)\tilde \epsilon_t-\tilde W_\text{inf}(t)\right),
\end{equation}  
where $f(\tilde \epsilon)=1/[1+\exp(\tilde \epsilon)]$, $\tilde \epsilon_t=\epsilon_t/(k_BT)$ and the (dimensionless) work for an infinite size reservoir $\tilde W_\text{inf}(t)$ given by Eq. (\ref{Winf}). The total work $\tilde W(t)=W(t)/(k_BT)$ can similarily be written, from Eq. (\ref{eq:extractedworkpert})
\begin{equation}
\tilde W(t)=\tilde W_\text{inf}(t)+  \frac{1}{\tilde C_T}\delta \tilde W(t), \hspace{0.5 cm} \delta \tilde W(t)=\frac{1}{2}\left\{\left[\tilde \epsilon_tf(\tilde \epsilon_t)-\tilde W_\text{inf}(t)\right]^2-\tilde \epsilon_0^2f^2(\tilde \epsilon_0)\right\},
\end{equation}  
and hence the heat
\begin{equation}
\tilde Q(t)=\tilde Q_\text{inf}(t)+ \frac{1}{\tilde C_T}\delta \tilde Q(t), \hspace{0.5 cm} \tilde Q_\text{inf}(t)=-\tilde W_\text{inf}(t)+f(\tilde \epsilon_t) \epsilon_t-f(\tilde \epsilon_0) \epsilon_0,  \hspace{0.5 cm} \delta \tilde Q(t)=-\delta \tilde W(t)+\delta p_1(t)\tilde \epsilon_t-\delta p(0)\tilde \epsilon_0,
\end{equation} 
where $\tilde Q(t)=Q(t)/(k_BT)$. Along the same lines one can expand the system entropy change, giving
\begin{equation}
\Delta S_{\rm S}(t)=\Delta S_\text{inf}(t)+ \frac{1}{\tilde C_T}\delta \Delta S(t), 
\end{equation}  
with
\begin{equation}
\Delta S_\text{inf}(t)/k_B=f(\epsilon_0)\ln\left[f(\epsilon_0)\right]+[1-f(\epsilon_0)]\ln[1-f(\epsilon_0)]-\left\{f(\epsilon_t)\ln\left[f(\epsilon_t)\right]+[1-f(\epsilon_t)]\ln[1-f(\epsilon_t)]\right\},
\end{equation}  
and
\begin{equation}
\delta \Delta S(t)/k_B=\delta p_1(t)\tilde \epsilon_t-\delta p(0)\tilde \epsilon_0.
\end{equation} 
We can thus write the first entropy
\begin{equation}
\Sigma_1(t)/k_B=\Delta S_\text{inf}(t)/k_B - \tilde Q_\text{inf}(t)+\frac{1}{\tilde C_T}\left[\delta \Delta S(t)/k_B-\delta \tilde Q(t)\right]=\frac{1}{\tilde C_T}\delta \tilde W(t),
\end{equation} 
where we used that $\Delta S_\text{inf}(t)/k_B=\tilde Q_\text{inf}(t)$. We note that we can write
\begin{equation}
\delta \tilde W(t)=\frac{1}{2}\left\{\left[\tilde Q_\text{inf}(t)+\tilde \epsilon_0f(\tilde \epsilon_0)\right]^2-\tilde \epsilon_0^2f^2(\tilde \epsilon_0)\right\},
\end{equation} 
which has no definite sign. However, as pointed out above, the expression for $\Sigma_1(t)$ is derived under the assumption that there are no initial correlations between the dot and reservoir state. Here we consider a fixed initial total energy $\mathcal{E}(0)$. This implies that the absence of initial correlations is only ensured for $\tilde \epsilon_0=0$, since the initial bath energy otherwise depends on the initial dot occupation.
In this case 
\begin{equation}
\Sigma_1(t)/k_B=\frac{1}{\tilde C_T}\frac{\tilde Q_\text{inf}^2(t)}{2} \geq 0,
\end{equation} 
in agreement with the second law. To get an expression for the second entropy $\Sigma_2(t)$ we first expand 
\begin{equation}
C'\left(T[Q(t)]-T]\right) \approx \tilde Q(t)+\frac{1}{\tilde C_T}\frac{\tilde Q^2(t)}{2}\approx \tilde Q_\text{inf}(t)+\frac{1}{\tilde C_T}\left[\frac{\tilde Q_\text{inf}^2(t)}{2}+\delta \tilde Q(t)\right].
\end{equation} 
We can then write
\begin{equation}
\Sigma_2(t)/k_B=\Delta S_\text{inf}(t)/k_B - \tilde Q_\text{inf}(t)+\frac{1}{\tilde C_T}\left[\delta \Delta S(t)/k_B-\delta \tilde Q(t)-\frac{\tilde Q_\text{inf}^2(t)}{2}\right]=\frac{1}{\tilde C_T}\left(\delta \tilde W(t)-\frac{\tilde Q_\text{inf}^2(t)}{2}\right).
\end{equation} 
Along the same line as for $\Sigma_1$, the assumption of no initial system-reservoir correlations then gives $\delta \tilde W(t)=\tilde Q_\text{inf}^2(t)/2$ and hence
\begin{equation}
\Sigma_2(t)=0 \leq \Sigma_1(t),
\end{equation} 
in line with both the second law and the relation in Eq. (\ref{eq:enthyr}).

\end{document}